\documentclass{iaus}
\usepackage{graphicx}

\title[Spectral Classification and Stellar Properties from Galactic Surveys] 
{Toward constraints on galaxy formation scenarios: stellar properties from Galactic surveys}

\author[Re Fiorentin, Bailer-Jones, Beers, Zwitter, Lee,  \& Xue]
{Paola Re Fiorentin$^{1,2}$, 
Coryn A. L. Bailer-Jones$^2$,
Timothy C. Beers$^3$,
Toma$\rm{\check{z}}$ Zwitter$^1$,
Young Sun Lee$^3$
 \and Xiangxiang Xue$^2$}

\affiliation{
$^1$Department of Mathematics and Physics, University of Ljubljana, \\ Jadranska 19,
SLO-1000 Ljubljana, Slovenia \\ email: {\tt paola.refiorentin@fmf.uni-lj.si, tomaz.zwitter@fmf.uni-lj.si} \\[\affilskip]
$^2$Max-Planck-Institut f{\"u}r Astronomie, \\ K{\"o}nigstuhl 17, D-68117 Heidelberg, Germany \\email: {\tt calj@mpia.de, xue@mpia.de}  \\[\affilskip]
$^3$Department of Physics \& Astronomy, CSCE: Center for the Study of Cosmic Evolution and JINA: Joint Institute for Nuclear Astrophysics, Michigan State University, \\ East Lansing, MI 48824, USA \\email: {\tt beers@pa.msu.edu, lee@pa.msu.edu}}

\pubyear{2008}
\volume{254}  
\pagerange{101--108}
\jname{The Galaxy disk in a cosmological context}
\editors{J. Andersen, J. Bland-Hawthorn \& B. Nordstr\"om, eds.}

\begin{document}

\maketitle

\begin{abstract}
We present some of the strategies being developed to classify and parameterize objects obtained with spectra from the Sloan Digital Sky Survey (SDSS) and the RAdial Velocity Experiment (RAVE) and present some results.  We estimate stellar atmospheric parameters (effective temperature, gravity, and metallicity) from spectral and photometric data and use these to analyse Galactic populations. We demonstrate this through the selection of a sample of candidate Blue Horizontal-Branch and RR Lyrae stars selected from SDSS/SEGUE.

\end{abstract}

\firstsection 
\section{Introduction}

The nature of the stellar populations of the Milky Way galaxy remains an
important issue for astrophysics, because it addresses the question of galaxy
formation and evolution and the origin of the chemical elements. To date,
however, studies have been limited by the small number of stars that could
be confidently identified as members of the various populations, and also by the
lack of available spectroscopy from which radial velocities and estimates of
atmospheric parameters (such as effective temperature, surface gravity, and
metallicity) could be obtained. 

With new ground-based and space-born survey missions currently under way or on
the immediate horizon, such as SDSS, RAVE, LAMOST, and Gaia, we are in the
golden age of Galactic astronomy. However, the classification of such a wide
variety of objects coming available is a challenging one, which requires
appropriate automated multi-dimensional data analysis techniques, and is a
necessary step toward constraining scenarios of Galaxy formation.

\section{Data}

As training grounds and complements to Gaia, we here focus on the analysis of data coming available from two complementary on-going spectroscopic surveys.

{\underline{\it The Sloan Digital Sky Survey}} 

In the northern hemisphere, SDSS-I and its extension for Galactic studies,
SDSS-II/SEGUE, have provided $9500$ square degrees of imaging data (position and
multicolor photometry) for over $200$ million stars, taken with a dedicated
2.5-m telescope on Apache Point, New Mexico. Some $3500$ square degrees at lower
Galactic latitudes ($|b| < 40^o$) are included as well.  


In addition to the imaging, these surveys obtained stellar spectra, covering the
wavelength range $\lambda\lambda$ 3850--9000\,$\AA$ at a resolving power $R
\simeq 2000$, for approximately $300\,000$ Galactic stars in the magnitude range
$14.0 \le g \le 20.5$; the radial velocities have typical accuracy of $10\,{\rm km\,
s^{-1}}$ (e.g., \cite[Adelman-McCarthy et al. 2008]{sdss_dr6}; \cite[Beers et al.
2004]{segue}).
 
{\underline{\it The RAdial Velocity Experiment}}

In the southern hemisphere, using the 6dF multi-object spectrograph on the 1.2-m
UK Schmidt Telescope of the Anglo-Australian Observatory, RAVE will measure
radial velocities and stellar atmospheric parameters of up to one million bright
stars by 2010. It has already observed over $250\,000$ stars away from the plane
of the Milky Way ($|b|>25^o$) in the magnitude range $9 < I < 12$, obtaining
medium-resolution spectra ($R \simeq 7500$) in the CaII triplet region
($\lambda\lambda$ 8410--8795\,$\AA$). 

In addition to cross-identification with
photometric and astrometric catalogues, the second data release provides
spectroscopic radial velocities with accuracy better than $2\,{\rm km\,s^{-1}}$
for about $50\,000$ stars, and stellar parameters for over $20\,000$ spectra
(\cite[Zwitter et al. 2008]{rave_dr2}).


\section{Spectral Analysis and Classification}

The main objectives of the classification are a discrete source classification
which might account for the identification of new types of objects and the
estimation of astrophysical parameters for specific classes. We use SDSS/SEGUE
and RAVE spectra to develop, implement, and test several methods for this
classification. 


{\underline{\it Principal Component Analysis}}

In terms of classification, a spectrum contains a large amount of redundant
information. We investigate the application of principal components analysis
(PCA) to the optimal compression of spectra.

\begin{figure}[h]
\begin{center}
 \includegraphics[angle=0,width=1\linewidth]{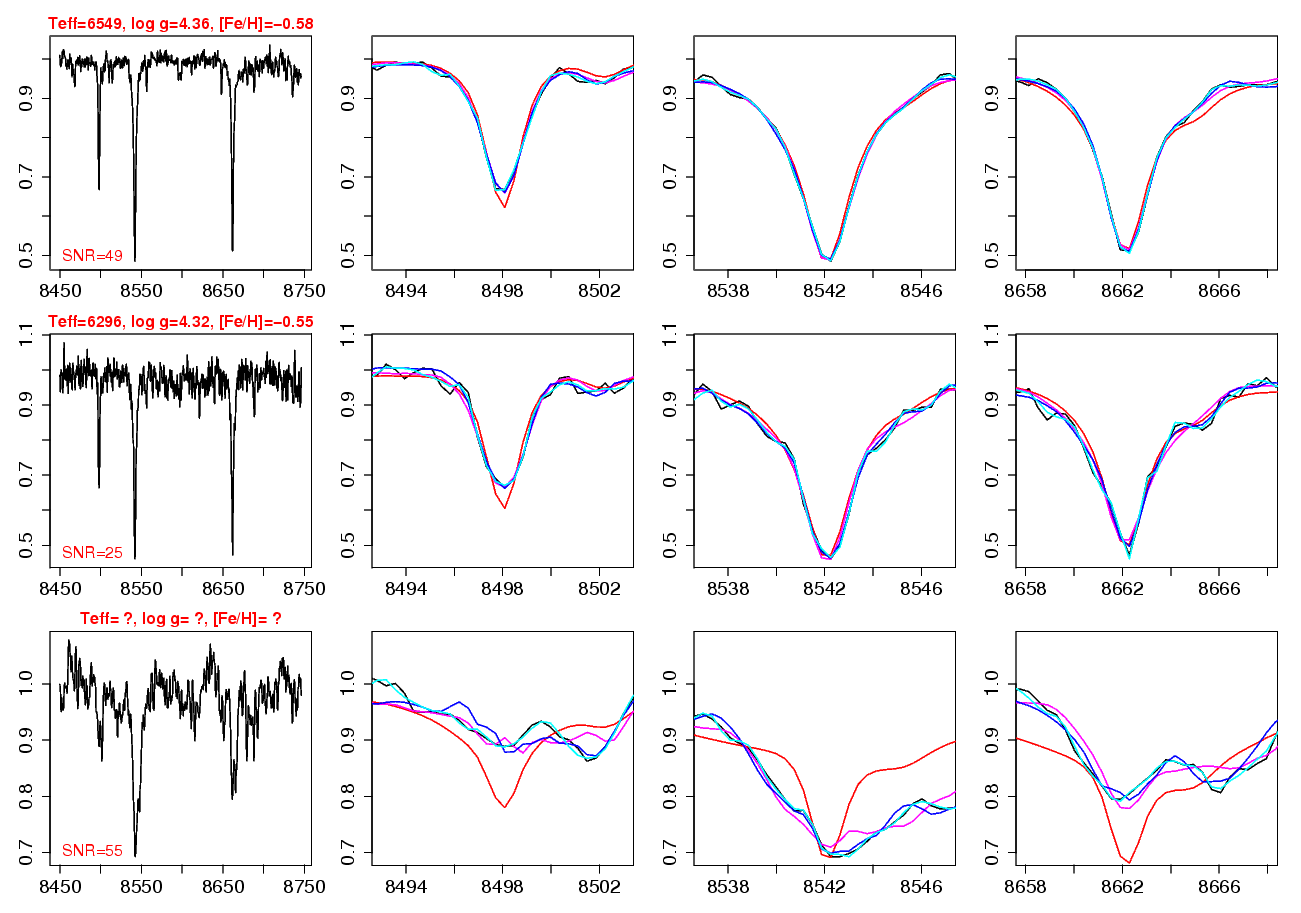} 
 \caption{Reconstruction of RAVE spectra (black lines) by projection onto synthetic principal components.
In each row, the spectrum on the left is the original and the following show, zooming in around
the CaII triplet, its reconstructions using 5 (red), 50 (pink), 100 (blue), and
250 (cyan) principal components.} \label{fig2}
\end{center}
\end{figure}

Using a sample of synthetic spectra (\cite[Munari et al. 2005]{munari}), we use
PCA to form a set of linearly independent basis vectors with which to describe
the data themselves, as well as any other newly observed spectrum.
Figure\,\ref{fig2} shows reduced spectral reconstructions (coloured lines)
around the CaII triplet for three selected RAVE spectra
(black lines), using different numbers of principal components computed from the
synthetic spectra.  The top and middle rows refer to the spectra of single stars
with similar atmospheric parameters, as obtained previously and listed in the
current catalogue, but with different of signal-to-noise ratios (namely,
$SNR=49$ and $25$ respectively).  The bottom row refers to the spectrum of a binary with
high signal-to-noise ($SNR=55$).

From inspection of these reconstructions, one can see how the PCA approach -- by
keeping only the most significant few components -- is able to retain essentially
all of the useful information and, beside being able to recover missing and/or borderline
features, acts as an effective filter to remove noise in the single star data
set (see Fig.\,\ref{fig2}, top and middle rows). Furthermore, it is shown that this
compression, which optimally removes noise, is able to isolate rare types of stars
with strong features such as binaries (see Fig.\,\ref{fig2}, bottom row).
As the template library only covers regular stars, peculiar objects (e.g.,
double-lined spectroscopic binaries and emission line objects) and outliers
(e.g., new types of objects) are not represented correctly with a few principal components, which allows them to be efficiently detected. Thus, beside
computational reasons (robustness and speed) and the higher accuracy achieved, this
method can be used to identify/classify unusual spectra and discover natural
classes among the data.

{\underline{\it Discrete Source Classifier}} 

In order to classify all sources (determining whether an object is a single star, binary, etc.), the DSC uses as its input spectra compressed via PCA (for speed and robustness of the classifier). 

The algorithm is trained on synthetic spectra.
Eight classes of astrophysical objects are considered.  In addition to regular spectra from the stellar library by \cite[Munari et al. (2005)]{munari}, libraries for peculiar objects are built (Re Fiorentin et al. in preparation) and adopted. The former sample provides cool (T1: $T_{\rm eff}< 5000$~K), medium-temperature (T2: $5000$~K $ \le T_{\rm eff}< 10000$~K), hot (T3: $T_{\rm eff} \ge 10000$~K), single stars with different levels of rotation, fast (F: $V_{\rm rot} \ge 50$~${\rm km\, s^{-1}}$) or low (L: $V_{\rm rot}< 50$~${\rm km\,s^{-1}}$); the latter includes binaries (B) and emission-core (EC) spectra.


To test the classification algorithms, samples of sources are selected both from
the synthetic data (SS) grids, with half of the original sample randomly
selected and not part of the training phase, and from observed RAVE data (SR)
that have been pre-classified via visual inspection. 

Algorithms implemented and tested are k-Nearest Neighbour (KNN; classifies
objects according by a majority vote of their neighbours), Artificial Neural
Networks (ANN), Support Vector Machines (SVM; classify the data by projecting the input space into a higher dimensional space and there finding optimal linear
discriminants between the classes). The interested reader is referred to Hastie
et al. (2001) for details.

Since the astrophysical classes of the test set are in fact known, the
statistical performance of the classifier can be assessed.  The Confusion
Matrices for the KNN, ANN, SVM classifiers are given in the tables. Results
refer to the evaluation of the synthetic sample (see Table \ref{tab1}) and a
selected RAVE sample (see Table \ref{tab2}), after training on the synthetic
spectra. Rows correspond to the true class of the test objects, and columns show
the classification results as a percentage of the total input sources of that
class. The leading diagonal indicates sources that are correctly classified, off
diagonal elements show the misclassification rates. 

\begin{table}
\begin{center}
  \caption{ Confusion Matrices for the KNN, ANN, SVM classifiers. Results refer to the evaluation synthetic sample after training on synthetic spectra (SS).}
  \label{tab1}
 {\scriptsize
        \begin{tabular}{ll|rrrrrrrr}\hline
        Method & true class &T1L & T1F & T2L & T2F & T3L & T3F & B & EC  \\
        \hline
       KNN & T1L & $98.06$ & $0.04$ & $1.60$ & $0.00$ & $0.00$ & $0.00$ & $0.30$ & $0.00$\\
& T1F & $0.08$ & $95.11$ & $0.00$ & $4.71$ & $0.00$ & $0.00$ & $0.00$ & $0.08$\\
& T2L & $0.46$ & $0.00$ & $99.17$ & $0.12$ & $0.01$ & $0.00$ & $0.24$ & $0.00$\\
& T2F & $0.00$ & $0.51$ & $0.15$ & $99.17$ & $0.00$ & $0.15$ & $0.00$& $0.01$ \\
& T3L & $0.00$ & $0.00$ & $0.39$ & $0.00$ & $98.55$ & $1.06$ & $0.00$ & $0.00$\\
& T3F & $0.00$ & $0.00$ & $0.00$ & $1.29$ & $4.29$ & $94.42$ & $0.00$ & $0.00$\\
& B   & $0.80$ & $0.04$ & $1.66$ & $0.32$ & $0.00$ &$0.00$ & $97.18$ & $0.00$\\
& EC  & $0.00$ & $1.48$ & $0.89$ & $1.78$ & $0.00$ & $0.00$ & $0.00$ &$95.85$\\ 
 \hline
        ANN &T1L & $92.61$ & $0.08$ & $7.17$ & $0.12$ & $0.00$ & $0.00$ & $0.01$ & $0.00$\\
& T1F & $0.36$ & $70.22$ & $0.00$ & $28.53$ & $0.00$ & $0.00$ & $0.00$ & $0.00$\\
& T2L & $2.34$ & $0.00$ & $95.38$ & $1.67$ & $0.00$ & $0.00$ & $0.60$ & $0.00$\\
& T2F & $0.00$ & $0.29$ & $1.68$ & $97.92$ & $0.00$ & $0.04$ & $0.07$& $0.00$ \\
& T3L & $0.00$ & $0.00$ & $10.45$ & $18.55$ & $0.23$ & $70.84$ & $0.00$ & $0.00$\\
& T3F & $0.00$ & $0.00$ & $0.00$ & $35.29$ & $0.00$ & $64.71$ & $0.00$ & $0.00$\\
& B   & $5.34$ & $0.46$ & $11.38$ & $3.74$ & $0.08$ &$0.10$ & $78.92$ & $0.00$\\
& EC  & $1.19$ & $0.89$ & $0.00$ & $12.76$ & $0.00$ & $0.89$ & $1.48$ &$82.79$\\
       \hline
        SVM & T1L & $99.62$ & $0.00$ & $0.04$ & $0.00$ & $0.00$ & $0.00$ & $0.34$ & $0.00$\\
& T1F & $0.00$ & $99.02$ & $0.00$ & $0.44$ & $0.00$ & $0.00$ & $0.53$ & $0.08$\\
& T2L & $0.00$ & $0.00$ & $100.00$ & $0.00$ & $0.00$ & $0.00$ & $0.00$ & $0.00$\\
& T2F & $0.00$ & $0.07$ & $0.00$ & $99.91$ & $0.00$ & $0.02$ & $0.00$& $0.00$ \\
& T3L & $0.00$ & $0.00$ & $0.00$ & $0.00$ & $100.00$ & $0.00$ & $0.00$ & $0.00$\\
& T3F & $0.00$ & $0.00$ & $0.00$ & $0.00$ & $0.53$ & $99.47$ & $0.00$ & $0.00$\\
& B   & $0.36$ & $0.06$ & $1.66$ & $0.04$ & $0.14$ &$0.00$ & $99.40$ & $0.00$\\
& EC  & $0.00$ & $0.00$ & $0.89$ & $1.48$ & $0.00$ & $0.00$ & $0.00$ &$97.63$\\
       \hline
       \end{tabular}
}
  \end{center}
\vspace{1mm}
 \scriptsize{
 {\it Class description (see discussion in text):}\\
  T1L: cool low rotating single star \\
  T1F: cool fast rotating single star\\ 
  T2L: medium-temperature low rotating single (normal) star \\
  T2F: medium temperature fast rotating single star\\ 
  T3L: hot low rotating single star \\
  T3F: hot fast rotating single star\\ 
  B: binary star \\
  EC: emission-core star\\ 
}
\end{table}

\begin{table}
\begin{center}
  \caption{ Confusion Matrices for the KNN, ANN, SVM classifiers. Results refer to the evaluation observed RAVE sample after training on synthetic spectra (SR). Classes are labeled as described in  Table \ref{tab1}.}
  \label{tab2}
 {\scriptsize
        \begin{tabular}{ll|rrrrrrrr}\hline
        Method & true class &T1L & T1F & T2L & T2F & T3L & T3F & B & EC  \\
        \hline
	KNN&T1L & $-$ & $-$ & $-$ & $-$ & $-$ & $-$ & $-$ & $-$\\
& T1F & $-$ & $-$ & $-$ & $-$ & $-$ & $-$ & $-$ & $-$\\
& T2L & $0.52$ & $0.00$ & $98.96$ & $0.16$ & $0.00$ & $0.00$ & $0.00$ & $0.36$\\
& T2F & $0.00$ & $0.00$ & $9.98$ & $87.72$ & $0.00$ & $0.19$ & $1.73$& $0.38$ \\
& T3L & $-$ & $-$ & $-$ & $-$ & $-$ & $-$ & $-$ & $-$\\
& T3F & $-$ & $-$ & $-$ & $-$ & $-$ & $-$ & $-$ & $-$\\
& B   & $4.58$ & $3.05$ & $14.50$ & $39.69$ & $0.00$ &$0.00$ & $37.40$ & $0.76$\\
& EC  & $2.72$ & $00.74$ & $4.44$ & $24.44$ & $0.00$ & $0.00$ & $21.73$ &$59.26$\\
	\hline
        ANN & T1L & $-$ & $-$ & $-$ & $-$ & $-$ & $-$ & $-$ & $-$\\
& T1F & $-$ & $-$ & $-$ & $-$ & $-$ & $-$ & $-$ & $-$\\
& T2L & $0.52$ & $0.00$ & $98.96$ & $0.16$ & $0.00$ & $0.00$ & $0.00$ & $0.36$\\
& T2F & $0.00$ & $0.00$ & $9.98$ & $87.72$ & $0.00$ & $0.19$ & $1.73$& $0.38$ \\
& T3L & $-$ & $-$ & $-$ & $-$ & $-$ & $-$ & $-$ & $-$\\
& T3F & $-$ & $-$ & $-$ & $-$ & $-$ & $-$ & $-$ & $-$\\
& B   & $4.58$ & $3.05$ & $14.50$ & $39.69$ & $0.00$ &$0.00$ & $37.40$ & $0.76$\\
& EC  & $2.72$ & $00.74$ & $4.44$ & $24.44$ & $0.00$ & $0.00$ & $21.73$ &$59.26$\\
        \hline 
	SVM & T1L & $-$ & $-$ & $-$ & $-$ & $-$ & $-$ & $-$ & $-$\\
& T1F & $-$ & $-$ & $-$ & $-$ & $-$ & $-$ & $-$ & $-$\\
& T2L & $0.52$ & $0.00$ & $98.96$ & $0.16$ & $0.00$ & $0.00$ & $0.00$ & $0.36$\\
& T2F & $0.00$ & $0.00$ & $9.98$ & $87.72$ & $0.00$ & $0.19$ & $1.73$& $0.38$ \\
& T3L & $-$ & $-$ & $-$ & $-$ & $-$ & $-$ & $-$ & $-$\\
& T3F & $-$ & $-$ & $-$ & $-$ & $-$ & $-$ & $-$ & $-$\\
& B   & $4.58$ & $3.05$ & $14.50$ & $39.69$ & $0.00$ &$0.00$ & $37.40$ & $0.76$\\
& EC  & $2.72$ & $00.74$ & $4.44$ & $24.44$ & $0.00$ & $0.00$ & $21.73$ &$59.26$\\
       \hline
       \end{tabular}
}
  \end{center}
\end{table}

Beside the excellent results obtained in the SS approach, those obtained for real
spectra (SR) appear promising. For all methods, the greatest confusion occurs
between binaries and emission-core stars. However, as they define a common
category of problematic objects, their peculiarity is highlighted and further
improvement may be needed to develop a specific classifier. In this approach,
calibration between the synthetic and observed data, and modeling of the noise (which
acts as regularizer) are required, and currently under study.

The KNN method has difficulties either with accuracy, which is degraded by the
presence of irrelevant features, and noise, or with computational efficiency in
high multidimensional data space. The early results from the ANN are promising,
and may be developed further, as some fine tuning is needed.  The SVM
method appears to be robust and reliable. However, at this stage, we can certainly
benefit from putting together the results obtained with these three classifiers
in a coherent picture.


{\underline{\it Stellar Atmospheric Parameters}} 

Once objects of interest are identified among all those observed, we focus
on the determination of their fundamental stellar atmospheric parameters.
Doing this we can consider not only normal (i.e., medium-temperature
and low rotating) single stars, but other classes of stars as well (although
this step has yet to be accomplished).

\begin{figure}[b]
\begin{center}
 \includegraphics[angle=-90,width=1\linewidth]{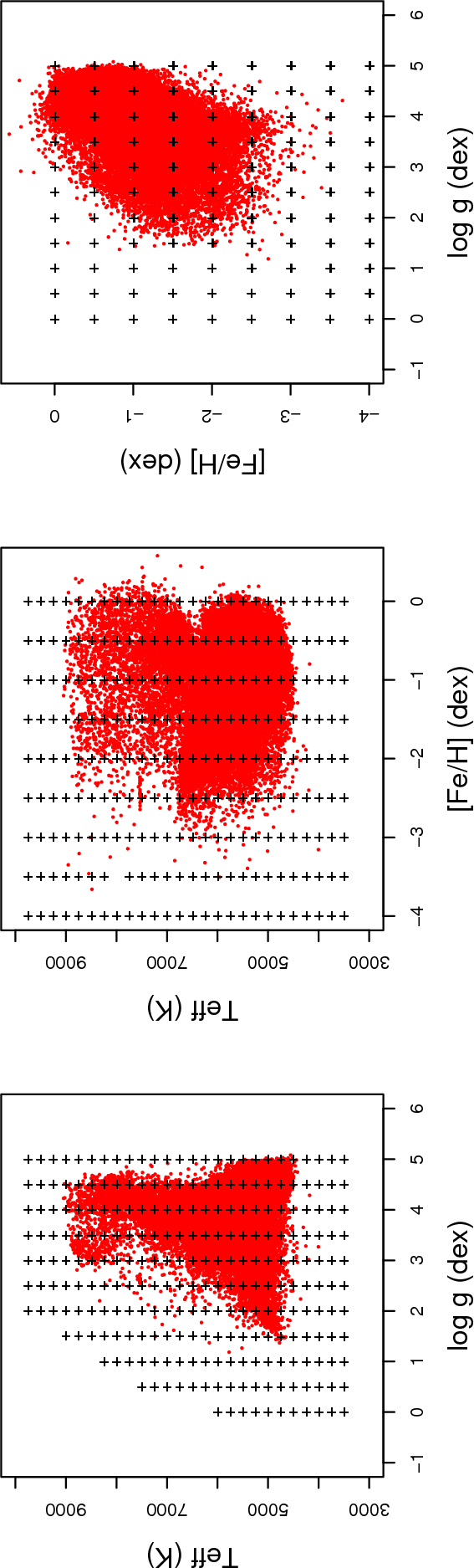} 
 \caption{The grid of stellar atmospheric parameters  $T_{\rm eff}$,
     $\log~g$, and ${\rm [Fe/H]}$. The synthetic parameters (plus symbols) are presented in comparison with previously estimated atmospheric parameters (dots) for 61\,069 SDSS/SEGUE spectra.}
   \label{fig3}
\end{center}
\end{figure}


In what follows, we focus on (normal) stars from the SDSS/SEGUE survey. From the
observed stellar spectra, we derived models to estimate effective temperature,
surface gravity, and metallicity via non linear regression models; these methods
have been described in detail by \cite[Re Fiorentin et al. (2007)]{prf07}, to
which the interested reader is referred for more details. 

Basically, an ANN is used for parametrization by giving a functional mapping
between PCA pre-processed stellar spectra as its inputs and the parameters
at its outputs. Optimal mapping is achieved by training 
on a set of pre-classified observed data (e.g., \cite[Lee et al. 2008]{sspp1})
and synthetic stellar templates derived from Kurucz's model atmospheres; the
grid of the available parameters is shown in Fig.\,\ref{fig3}. 

From independent subsamples not involved in the training phase, the accuracies
of our predictions (mean absolute errors) for each parameter are $T_{\rm eff}$
to $78$~K ($111$~K), $\log~g$ to $0.17$~dex ($0.31$~dex), and ${\rm [Fe/H]}$ to
$0.09$~dex ($0.24$~dex), respectively. 
The precisions achieved now are about $50\%$ better than those reported in
\cite[Re Fiorentin et al. (2007)]{prf07}, and are the result of further
development of the regression models, improved stellar models, and better data
calibrations. 

Stellar atmospheric parameters 
are then derived for 186\,580 stellar spectra (from SDSS/SEGUE plates) with
signal-to-noise ratio $SNR > 10$. This sample having, along with the 
photometry and radial velocities, 
is suitable to carry out Galactic studies; the following application is based on
such a dataset.

\section{Application: Stellar Properties of Galactic Tracers/Populations from SDSS}

Here we illustrate the capability of an efficient classification and parameter
estimation effort in the context of constraining Galaxy formation scenarios.
Therefore, aiming to investigate the presence of substructures, we focus on Blue
Horizontal-Branch (BHB) stars and RR Lyrae stars, which are excellent tracers to study
Galactic populations, as they are nearly standard candles, and sufficiently bright to
be detected up to large distances. 

\begin{figure}[b]
\begin{center}
 \includegraphics[angle=-90,width=0.5\linewidth]{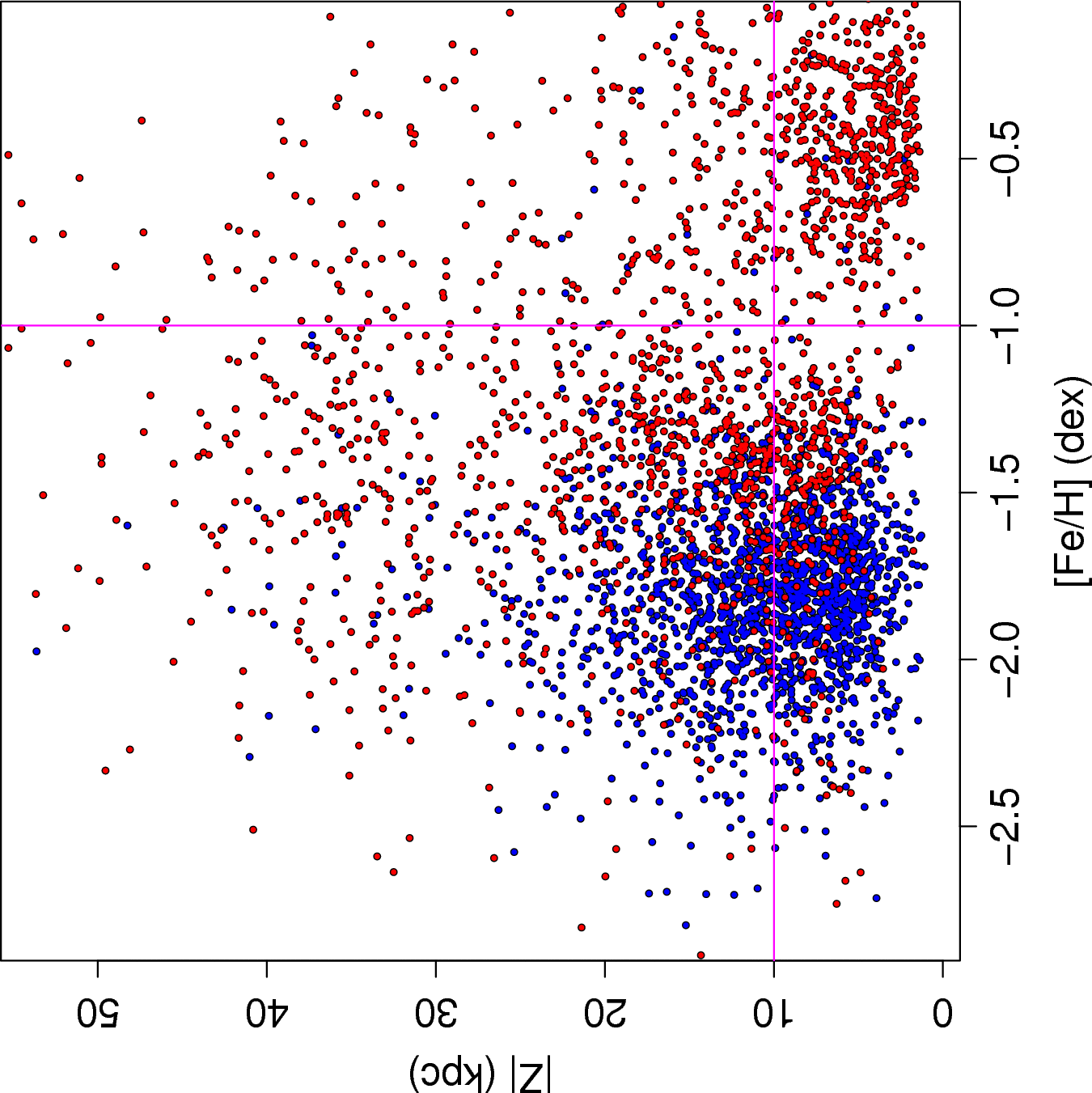} 
\caption{Metallicity distribution of the $4123$ stars selected as BHB (blue) or
RR Lyrae (red), as a function of distance from the Galactic plane. 
Lines show constraints on the subsequent tracers selection.}
   \label{fig4}
\end{center}
\end{figure}

{\underline{\it Candidate selection and distance estimates}}

In order to obtain pure tracers samples with no (or little) contamination, we combine simple colour cuts with the spectroscopic or atmospheric parameter information.

 
Due to multiple observations of the same stars in different spectrographic
plugplates, the spectro-photometric sample consists effectively of $168\,340$
unique objects for which the parameters assigned depend on the signal-to-noise level of the spectra.  

The $2517$ BHB stars of our sample were identified employing a stringent
approach which combines colour cuts previously established by \cite[Yanny et al.
(2000)]{yanny} with a set of Balmer-line profile selection criteria (see
\cite[Xue et al. 2008]{xue}; \cite[Sirko et al. 2004]{sirko}). These tracers
have the best constrained absolute magnitude ($M_g=0.7$) and thus allow the
derivation of accurate photometric distances. 

The $1606$ RR Lyrae stars of our sample were assembled from the colour
selection method proposed by \cite[Ivezi$\rm{\acute{c}}$ et al. (2005)]{ivezic}, adopting a completeness of $C=50\%$ and efficiency of $E=35\%$. This was then achieved by adopting the following cuts in parameter space: $6100$~K $<T_{\rm eff}< 7400$~K and $3$~dex $<\log~g< 4$~dex. 
The absolute magnitude of RR Lyrae stars correlate quite closely with metallicity ${\rm [Fe/H]}$; we adopt the empirical linear relation given by \cite[Kinman et al. (2007)]{kinman} and estimate their distances.




{\underline{\it Spatial and Metallicity Distribution}}

The observed magnitudes of our sample of $4123$ Galactic tracers have been used
to infer their distances and, consequently, their Galactic distribution.

Figure\,\ref{fig4} shows the metallicities and distances from the plane for BHB
stars (blue) and RR Lyrae stars (red). Globally, we see a gradient of ${\rm
[Fe/H]}$ with respect to $|Z|$. Inspection of this distribution shows the BHB
and RR Lyrae stars as tracers having different intrinsic physical properties:
while the former are essentially metal-poor halo stars that do not reach
to the farthest distances, the latter include halo stars 
and old disk stars 
that extend deep into the outer
halo. 
They represent the disk and halo populations which here
appear remarkably well-defined and separated; the stellar properties obtained
help describing such Galactic populations.

Furthermore, we can see possible clumps which suggest hints of substructures,
the possible fossil signatures of past merging events.  We are in the
process of quantifying these via tests of clustering
in spatial, metallicity, and radial velocity space.

\section{Summary}

We have implemented machine learning algorithms to classify observed objects and to
determine their atmospheric parameters. Here, discrete source
classifiers (from unsupervised and supervised analysis) are tested on RAVE
spectra, and results for parameter estimation of single stars are given for
SDSS/SEGUE spectra.

Based on the stellar atmospheric parameters which we have estimated from SDSS
spectra we can better select target objects (such as BHB and RR Lyrae stars) for
Galactic studies than by using photometry alone, and better explore the interface
between the thick-disk and halo populations.

The models are in the process of development to improve their accuracy,
and for the identification of peculiar/new types of objects, and their parameterization.

Looking further ahead, such strategies form the basis for future ground/space
astrometric missions classifiers that are essential, in particular,
for fully exploiting the astrometric part of the Gaia catalogue for stellar population studies. 

\begin{acknowledgments}
This work is supported through the Marie Curie Research Training Network ELSA
(European Leadership in Space Astrometry) under contract MRTN-CT-2006-033481
to P. Re~Fiorentin. 
\end{acknowledgments}

\end{document}